\documentclass[aps,prl,twocolumn,preprintnumbers]{revtex4}
\usepackage{graphicx}
\usepackage{dcolumn}
\usepackage{bm}
\def\be{\begin{equation}}
\def\ee{\end{equation}}
\def\bea{\begin{eqnarray}}
\def\eea{\end{eqnarray}}

\usepackage{latexsym}
\usepackage{epsfig,amssymb,euscript}
\usepackage{amsmath}
\usepackage{array,calc,epsfig}
\newcommand{\Tr}{{\rm Tr}}
\begin{document}
\title{Neutron electric dipole moment from gauge/string duality \\
} 

\author{Lorenzo Bartolini$\,^{a}$, Francesco Bigazzi$\,^{a,b}$, Stefano Bolognesi$\,^{b}$, Aldo L. Cotrone$\,^{a}$, Andrea Manenti$\,^{b,c}$}
\affiliation{\textit{a} Dipartimento di Fisica e Astronomia, Universit\`a di Firenze and INFN, Sezione di Firenze; Via G. Sansone 1, I-50019 Sesto Fiorentino (Firenze), Italy.\\
\textit{b} Dipartimento di Fisica ``E. Fermi", Universit\'a di Pisa and INFN, Sezione di Pisa; Largo Bruno Pontecorvo 3, I-56127 Pisa, Italy.\\
\textit{c} Institute of Physics, EPFL; Rte de la Sorge, BSP 728, CH-1015 Lausanne, Switzerland. $^*$}

\begin{abstract}
We compute the electric dipole moment of nucleons in the large $N_c$ QCD model by Witten, Sakai and Sugimoto with $N_f=2$ degenerate massive flavors. 
Baryons in the model are instantonic solitons of an effective five-dimensional action describing the whole tower of mesonic fields. 
We find that the dipole electromagnetic form factor of the nucleons, induced by a finite topological $\theta$ angle, exhibits complete vector meson dominance. We are able to evaluate the contribution of each vector meson to the final result - a small number of modes are relevant to obtain an accurate estimate.  Extrapolating the model parameters to real QCD data, the neutron electric dipole moment is evaluated to be $d_n = 1.8 \cdot 10^{-16}\, \theta\;e\cdot \mathrm{cm}$. The electric dipole moment of the proton is exactly the opposite.

\end{abstract}

\maketitle
Permanent electric dipole moments of fundamental or composite particles with spin are a sensible probe of CP-violating effects in nature.
In particular, measurements of the neutron electric dipole moment (NEDM) give the strictest bounds on the magnitude of the CP-violating $\theta$-term in QCD. 
The current bound, $|d_n| \leq 2.9 \times 10^{-26}\, e\cdot \mathrm{cm}$ \cite{pendlebury}, points towards an extremely small value of $\theta \sim {\cal O}(10^{-10})$, giving rise to the strong CP problem.

As usual in QCD, first principle calculations of the NEDM are extremely complicated due to the large coupling regime.
Estimates can be given with lattice techniques (which, in general, suffer from a sign problem at finite $\theta$) or phenomenological settings, such as the chiral Lagrangian and the Skyrme model. Writing $d_n = c_n\, \theta\cdot 10^{-16}\, e\cdot \mathrm{cm}$, these estimates give ${\cal O}(1)\lesssim|c_n|\lesssim{\cal O}(10)$ with most of the results pointing towards $c_n>0$. A detailed review on the matter can be found in \cite{vicari}.  

In this paper we provide a first principle computation of the NEDM in the Witten-Sakai-Sugimoto (WSS) model \cite{witten,SS1}, the top-down holographic theory closest to QCD.
Its main advantage over other phenomenological models is that it incorporates automatically the whole tower of vector mesons. Actually the model exhibits complete vector dominance in the hadron electromagnetic form factors. We will show, for the first time, that this feature is exhibited also by the CP-breaking dipole term. Thus, we are able not only to provide a novel estimate of the value of the NEDM, which turns out to be of the same order of magnitude of other results, but also to account for the contribution of each vector meson. As we will show, the outcome is that considering just the first vector mode is not sufficient to give an accurate estimate of the full result; instead, including the first three modes leaves just a per cent error. 
\\ 
{\it 1. Holographic Mesons}.
The WSS model is based on a $D4-D8$ brane setup in type IIA string theory. 
In the limit where a simple dual holographic description can be given, the model reduces to a $3+1$ dimensional large $N_c$ $SU(N_c)$ gauge theory with $N_f$ massless quarks. 
In addition, it also contains a tower of massive adjoint matter fields whose mass scale is set by a dimensionful parameter denoted as $M_{KK}$ (which gives the scale of glueballs as well).
They arise as modes of a Kaluza-Klein reduction of the $4+1$ dimensional theory on $N_c$ $D4$-branes on a circle.
How much this spurious sector can be decoupled from the QCD-like one, depends on a 't Hooft-like coupling $\lambda$, setting the ratio between the confining $SU(N_c)$ string tension and $M_{KK}^2$. 
The model has a simple dual holographic description if $\lambda\gg1$, when the two sectors are not decoupled. 
Despite this feature, at low energies, the model shares with QCD all the expected features like confinement, chiral symmetry breaking and so on. Moreover, since it is embedded in a well defined string theory setup, the corrections to the leading $\lambda\gg1$, $N_c\gg1$ behavior are, in principle, under control.

The flavor sector in the model is described by the low energy modes of $N_f$ $D8$-branes. 
The backreaction of these branes on the background is weighed by a parameter $\epsilon_f =(\lambda^2/12\pi^3)(N_f/N_c)$. 
When $\epsilon_f\ll1$ the $D8$-branes can be treated as probes of the background sourced by the $D4$'s: this corresponds to the quenched approximation for the quarks in the model. 
In this limit, their effective action reduces to a $U(N_f)$ Yang-Mills theory with Chern-Simons terms on a curved space-time in five dimensions \cite{SS1}
\bea\label{actions}
&& S_{WSS}= S_{YM} + S_{CS}\,,\\
&& S_{YM} = -\kappa\int d^4x d z\Tr\left(\frac{h(z)}{2} \mathcal{F}_{\mu\nu}\mathcal{F}^{\mu\nu} + k(z)\mathcal{F}_{\mu z}\mathcal{F}^\mu_{\;\; z}\right)\nonumber\\
&& S_{CS} = \frac{N_c}{24\pi^2}\int  \Tr\left(\mathcal{A}\mathcal{F}^{ 2}- \frac{i}{2}\mathcal{A}^{ 3}\mathcal{F} - \frac{1}{10}\mathcal{A}^{ 5}\right)\,,
\nonumber
\eea
where (in units $M_{KK}=1$) $\kappa = \frac{\lambda N_c}{216\pi^3}, h(z) = (1+z^2)^{-1/3}, k(z) = (1+z^2)$, and we have omitted the wedge product symbol ``$\wedge$". The effective theory for mesons emerges by inserting into the action above the following expansions for the gauge field:
${\mathcal A}_{z} (x^{\mu},z) = \sum_{n=0}^{\infty} \varphi^{(n)}(x^{\mu})\phi_n(z), {\mathcal A}_{\mu}(x^{\mu},z) = \sum_{n=1}^{\infty} B_{\mu}^{(n)}(x^{\mu})\psi_n(z)$. The functions $\phi_n(z)$, $\psi_n(z)$ form complete sets normalized in such a way that the fields $\varphi^{(n)}$ and $B_{\mu}^{(n)}$ get canonical mass and kinetic terms in four dimensions. In particular each function $\psi_n(z)$ is an eigenfunction, with eigenvalue $\lambda_n$, of the equation
\be \label{eqforpsi}
-h(z)^{-1} \partial_z (k(z)\partial_z\psi_n(z))=\lambda_n \psi_n(z)\,,
\ee
while $\phi_n(z) = \lambda_n^{-1/2}\partial_z\psi_n(z)$ for $n>1$ and $\phi_0(z) = (\kappa\pi)^{-1/2}k(z)^{-1}$. 
The modes $B_{\mu}^{(n)}$ correspond to massive vector (for odd $n$) and axial vector (for even $n$) fields with masses $m_n^2 = \lambda_n M_{KK}^2$. 
For example $B_{\mu}^{(1)}$ and $B_{\mu}^{(2)}$ correspond to the $\rho$ and the $a_1$ meson respectively. The scalar modes $\varphi^{(n)}$ with $n\ge1$ are eaten by the $B_{\mu}^{(n)}$, while $\varphi^{(0)}$ corresponds to the pion. 
The remarkable feature of the WSS 5d effective action is thus the fact that it automatically includes, into a unifying picture, not only the low lying modes (with their chiral Lagrangian with Skyrme and WZW terms) but also the tower of massive meson fields. 
Moreover, all the parameters in the meson effective action are given in terms of the few bare parameters of the model, i.e. $N_c,N_f,M_{KK},\lambda$. 
In the following we will focus on the $N_f=2$ case. \\
{\it 2. Adding $\theta$ and quark mass terms}. 
The CP-breaking topological term in the WSS model is automatically included as it corresponds to the $\int C_1 {\rm Tr} F\wedge F$ Chern-Simons term in the $D4$-branes action. 
This allows to relate the field theory $\theta$ parameter with the integral of the Ramond-Ramond $C_1$ potential along the circle wrapped by the $D4$-branes \cite{Wittentheta}. 
Adding the $\theta$ term to the model effectively amounts on adding the term \cite{SS1},
\be\label{actiontheta}
S_{\theta}= -\frac{\chi_g}{2}\int d^4x \left[\theta + \int dz {\rm Tr} {\mathcal A}_z \right]^2
\ee
to the five dimensional effective action (\ref{actions}).
In the expression above, $\chi_g = \lambda^3/(3^{6}4\pi^6)$ is (in units $M_{KK}=1$) the topological susceptibility of the unflavored theory. 
The $\int dz {\rm Tr} {\mathcal A}_z$ term is related to the $\eta'$ meson and the action above provides its mass $m_{\eta'}^2 = (\pi N_f/2\kappa)\chi_g \equiv (2N_f/f_{\pi}^2)\chi_g$ according to the Witten-Veneziano relation (with pion decay constant $f^2_{\pi} = 4\kappa/\pi$).  
Notice that this implies that $(m_{\eta'}^2/M_{KK}^2) \sim \epsilon_f \ll1$ in the regime we are working with. 

In the WSS model the pion matrix is given by the path ordered holonomy matrix $U = {\mathcal P} {\rm exp} [-i \int dz {\mathcal A}_z]$. Thus, in analogy with the chiral Lagrangian approach, the suitable mass term we have to add to the 5d effective action in order to describe massive quarks is
\be\label{actionmass}
S_M = c \int d^4x\, \Tr\mathcal{P}\left[M e^{-i\int \mathcal{A}_z dz}+ \mathrm{c.c.}\right]\,,  
\ee
where $c$ is a constant and $M$ is the mass matrix. 
This action has a precise meaning in string theory \cite{AK, Hashimotomass}: it is the deformation due to open string worldsheet instantons stretching between the $D8$-branes. 
The Nambu-Goto part of the open string action is put on-shell and its exponentiation contributes to the constant $c$ and the mass terms. 
What remains is just the boundary interaction of the open string with the gauge fields on the $D8$-branes. 
Notice that, when $S_M$ is added together with $S_{\theta}$, a gauge shift $\delta_{\Lambda}\int dz {\rm Tr} {\mathcal A}_z =-\theta$ does not remove the $\theta$-dependence. 
Instead, this produces a phase shift $M\rightarrow Me^{i\theta/N_f}$ on the mass matrix, so that the physical topological parameter in the model turns out to be the combination $\bar\theta \equiv \theta + \arg \det M$, as expected from field theory.
In the following we will focus on the case of degenerate flavors $M_{ij} = m_q \delta_{ij}$, choosing $m_q$ to be real. 
Writing the pion matrix as $U = \exp \left(2i \pi^a(x^\mu) T^a/f_{\pi}\right)$ where $T^a$ are $U(N_f)$ generators normalized as $\Tr(T^aT^b)=\delta_{ab}/2$,
expanding $S_M$ around the vacuum solution ${\mathcal A}=\mathbf 0$ (before adding the $\theta$-term), we can read off the Gell-Mann-Oakes-Renner (GMOR) relation $f_{\pi}^2 m_{\pi}^2 = 4 c m_q$ which in turn allows us to relate $c$ with the VEV of the chiral condensate. Adding the term $S_M$ to the original WSS action is justified in the limit $m_q\ll M_{KK}$, or, in terms of the pion mass, in the limit $m_{\pi}\ll M_{KK}$.

Our convention for the $U(N_f)$ gauge field follows \cite{SS1}: setting $N_f=2$, ${\mathcal A} \equiv A + ({\widehat A}/2)\mathbf{1}$, where $A=A^a\tau^a/2$ is the $SU(2)$ component ($\tau^a$, $a=1,2,3$, being the Pauli matrices) and ${\widehat A}$ is the Abelian one. 
The field theory vacuum following from the total action $S_{WSS}+S_{\theta}+S_M$ corresponds to the pure gauge configuration 
\begin{equation}\label{vac}
A=\mathbf 0\,, \qquad \int {\widehat A}_z dz = - \theta\,.
\end{equation}
A detailed analysis of the vacuum structure, which turns out to be precisely the same as the one from the QCD chiral Lagrangian, will be given in \cite{noi}. 
\\
{\it 3. Holographic Baryons}. 
The baryon in the WSS model is identified with an instantonic soliton (localized in the four directions $x^{M}=(x^i, z)$) of the 5d action (\ref{actions}) \cite{SS-barioni}. 
The baryon number coincides with the instanton number. 
This description resembles the Skyrme description of  (large $N_c$) baryons as solitons of the chiral Lagrangian \cite{ANWSkyrme}. The one instanton solution for $N_f=2$ in the original WSS model is analytically known around $z=0$; the general solution is presented in \cite{SS-form}. In the former limit, the curvature of the 5d ambient space can be neglected, the (classical) solution is just a BPST instanton \cite{BPSTInst} which, due to the presence of the CS terms, also sources an electric potential
\bea \label{instanton}
&&A_M^{\mathrm{cl}} = -i (1-f(\xi)) g^{-1} \partial_M g\,, \qquad A_0^{\mathrm{cl}}=\widehat{A}_M=0\,,\nonumber \\
&&\widehat{A}^{\mathrm{cl}}_0 = \frac{N_c}{8\pi^2\kappa}\frac{1}{\xi^2}\left[1-\frac{\rho^4}{(\rho^2+\xi^2)^2}\right]\,,
\eea
where $\xi^2 \equiv (z-Z)^2 + |\vec{x}-\vec{X}|^2$ and 
\begin{equation}
f(\xi) = \frac{\xi^2}{\xi^2+\rho^2}\,, \quad g(x) = \frac{(z-Z)\mathbf 1 - i (\vec{x}-\vec{X})\cdot \vec{\tau}}{\xi}\,.
\end{equation}
The solution depends on a set of parameters: the instanton size $\rho$, the instanton center of mass position $X^{M}$ in the 4d Euclidean space, and 3 $SU(2)$  ``angles" related to the fact that the solution can be rotated by means of a global gauge transformation. 
Evaluating the action $S_{WSS}$ on the above solution, one finds that it is minimized for $\rho^2 =\rho_{cl}^2= (N_c/8\pi^2 \kappa)\sqrt{6/5}$, $Z = Z_{cl}=0$. 
The other parameters are genuine instanton moduli (in the calculation below we will set $\vec X =0$ without loss of generality). 
The instanton quantum mechanics is described by the Hamiltonian for the instanton parameters, promoted to time (and space) dependent operators. \\
{\it 4. The NEDM}. 
The electric dipole moment of the neutron is defined as
\begin{equation}\label{nedm}
\vec{D}_{n,s} = \int d^3x\,\vec{x}\, \langle n,s | J^0_\mathrm{em} |n,s\rangle\,.
\end{equation}
Here $J_\mathrm{em}$ is the electromagnetic current and $|n,s\rangle$ is the neutron state with spin $s$.
Holographically, the state $|n,s\rangle$ has been derived in \cite{SS-barioni} in the zero mass, $\theta=0$ case. Omitting the spin-dependent part, $|n \rangle \propto P(\rho)\psi_Z(Z)$, with $P(\rho)=\rho^{-1+2\sqrt{1+N_c^2/5}}e^{-(8\pi^2\kappa\rho^2)/\sqrt{6}}$ and $\psi_Z(Z)=e^{-(8\pi^2\kappa Z^2)/\sqrt{6}}$. The NEDM is computed at leading order in $m_q,\theta$. In this regime, the neutron state is unchanged w.r.t. the one in \cite{SS-barioni}.
The electromagnetic current is extracted holographically from the 5d gauge field $\mathcal{F}$ as \cite{SS-form}
\begin{equation}\label{current}
\begin{aligned}
J_{\mu\,em}&=-\kappa\left[k(z)\Tr(F_{\mu z}\tau^3)+\frac{k(z)}{N_c}{\widehat F}_{\mu z}\right]^{z\to\infty}_{z\to-\infty}\,.
\end{aligned}
\end{equation}
The NEDM is computed by solving the equations of motion for $\mathcal{F}$ from the actions (\ref{actions}), (\ref{actiontheta}) and (\ref{actionmass}) and performing the integral in (\ref{nedm}) using the neutron state above.

The NEDM is extracted from the linear-in-$\theta$ term in the electromagnetic current.
We also work in the phenomenologically relevant case of small quark masses w.r.t. the dynamically generated scale of QCD.
Thus, the equations of motion are expanded at first order in $\theta$ and $m_q$, using as background the vacuum plus baryon solutions  (\ref{vac}), (\ref{instanton}). 
We also exploit the fact that close to $z \sim 0$ the time component of the gauge field $\mathcal{A}$ scales as $\lambda^0$ while the other components as $\lambda^{1/2}$ \cite{SS-form}; at large $\lambda$ this selects a few terms in the equations.
Moreover, at large $z$ all the background fields are power-like suppressed in $z$ and the equations are essentially linearized.

The only components of the bulk gauge field $\mathcal{A}$ relevant for the computation are $\widehat{A}_z$ and $A_0$. In particular, the ${\widehat F}_{0z}$ term from (\ref{current}) does not contribute to the dipole to leading order in $\theta$. The equations are then (in the conventions $\varepsilon_{0123z}=- \varepsilon^{0123z}=1$) 
\begin{eqnarray}
& -\kappa \left(h(z)D_\nu {F^{0\nu}} + D_z(k(z){F^{0 z}})\right)^a{\big|_\mathrm{mass}} \nonumber \\
& -\frac{N_c}{64\pi^2}\varepsilon^{ijk}\left(2 F_{ij}^a\widehat{F}^{\mathrm{mass}}_{kz} +  2 F_{iz}^a\widehat{F}^{\mathrm{mass}}_{jk}\right)=0 \label{eomA0}\,, \\
& -\kappa\, k(z)\partial_\nu\widehat{F}^{z\nu}_\mathrm{mass} =-\chi_g \left( \theta+ \int_{-\infty}^\infty {\widehat A}_z dz  \right) \nonumber \\
& - i c \, Tr \left[\frac{M}{2}\left(\mathcal{P}e^{-i\int_{-\infty}^\infty{\mathcal A}_z dz}- \mathrm{c.c.}\right)\right] \label{eomAz}\,.
\end{eqnarray}
With ``mass'' we denote the linear term in $m_q$. We use the static gauge and consider time independent gauge fields (time dependence gives subleading contributions \cite{SS-barioni,SS-form}).

Evaluating the various contributions, the equation of motion (\ref{eomAz}) for $\widehat{A}_z^\mathrm{mass}$ reads
\begin{equation} 
\kappa\, k(z)\partial_i\partial^i\widehat{A}_z^\mathrm{mass}  = c m_q \theta \left[\cos{\left( \frac{\pi}{\sqrt{1+\rho^2/r^2}}\right)} +  1 \right] \,,
\end{equation}
where $r^2=\vec x \cdot \vec x$.
Defining
$\widehat{A}^\mathrm{mass}_z \equiv \frac{u(r)}{1+z^2}$
gives the ODE
\begin{equation}
\frac{1}{r^2}\partial_r(r^2 \partial_r u(r)) = \frac{ c m_q }{\kappa}  \theta \left[\cos{\left( \frac{\pi}{\sqrt{1+\rho^2/r^2}}\right)} +  1 \right] \label{eqforu} \ .
\end{equation}

The function $u(r)$ enters as source in equation (\ref{eomA0}) for $A_0^{\rm mass}$.
The ansatz
$A_0^{\rm mass} = W(r,z)(\vec{x}-\vec{X})\cdot\vec{\tau}$
gives the equation
\begin{eqnarray}\label{eqforW}
& h(z) \left(\partial_r^2 W(r,z) + \frac{4}{r}\partial_r W(r,z) + \frac{8\rho^2}{(\xi^2+\rho^2)^2}W(r,z)\right) + \nonumber \\ 
& \partial_z(k(z) \partial_zW(r,z)) = \frac{27\pi}{\lambda} \frac{\rho^2}{(\xi^2+\rho^2)^2}\frac{1}{r}\frac{u'(r)}{1+z^2} \equiv {\cal G}(r,z)\,.
\end{eqnarray}
The role of the vector modes is extracted with the expansion
$W(r,z) = \sum_{n=1}^\infty R_n(r) \psi_n(z)$ where the $\rho,Z$ dependence is implicit.
Since the complete set of eigenfunctions $\psi_n(z)$ satisfies equation (\ref{eqforpsi}) and
$\kappa  \int dz \,h(z) \psi_n(z) \psi_m(z) = \delta_{mn}$,
we can project equation (\ref{eqforW}) 
on the $\psi_n(z)$, 
obtaining an infinite set of coupled equations of the form
\begin{eqnarray}
& \partial_r^2 R_m(r) + \frac{4}{r}\partial_r R_m(r) - \lambda_m R_m(r) + \\
&\sum_{n=1}^\infty \left\langle m \left| \frac{8\rho^2}{(\xi^2+\rho^2)^2}\right|n\right\rangle R_n(r) = \langle m | h^{-1}\, {\cal G} \rangle  \label{eqproj}\,, \nonumber
\end{eqnarray}
where
\begin{eqnarray}\label{matrices}
& \left\langle m \left| \frac{8\rho^2}{(\xi^2+\rho^2)^2}\right|n\right\rangle \equiv \kappa \int dz\, h(z) \psi_n(z)\psi_m(z) \frac{8\rho^2}{(\xi^2+\rho^2)^2}\,, \nonumber \\
& \langle m | h^{-1}\,{\cal G}\rangle \equiv \kappa\int dz\, \psi_m(z) {\cal G}(r,z)\,.
\end{eqnarray}
The system can be solved by truncating it at the desired value of $m$, once (\ref{eqforpsi}) and (\ref{eqforu}) have been solved (we do it numerically).

The electromagnetic current is now computed by switching on the gauge group orientation moduli $\vec{\bf a}$.
For the NEDM, it is sufficient to rotate $A_0 \longmapsto A_0' = V A_0 V^{-1}$ with $V$ a group valued function such that $V \rightarrow \vec{\bf a}$ as $z\rightarrow \pm \infty$ \cite{SS-form}. The $A_M$ components are also transformed accordingly. 
The rotated fields still satisfy the equations of motion.
The current (\ref{current}) is thus
\begin{equation}
J_{em}^0 = \kappa\left[k(z)V (\partial_z \Tr (A^0_\mathrm{mass}\tau^3))V^{-1}\right]^{z\to \infty}_{ z\to -\infty}\,,
\end{equation}
since the direct contribution of $\widehat{A}_z$ vanishes.
Using the identity for the generic baryonic state $B$
\begin{equation}
\langle B',s' | \Tr(\vec{\bf a} \tau^i \vec{\bf a}^{-1} \tau^a)|B,s\rangle = -\frac{2}{3}(\sigma^i)_{s's}(\tau^3)_{I_3'I_3}\,,
\end{equation}
with $\sigma$ and $\tau$ being the Pauli matrices for spin and isospin respectively (see e.g. \cite{ANWSkyrme}), the ``semi-classical'' part of the NEDM (i.e. the result before including the $\rho,Z$-dependent part of the neutron wave function) is given by $\vec{\mathcal{D}}^{s.c.}_{n,s}=d^{s.c}_n\langle s |\vec{\sigma}|s\rangle$, with 
\begin{equation}\label{nedmfinal}
d^{s.c.}_n = -\frac{8\pi}{9} \sum_{n=1}^{\infty} g_{v^n} \int_0^\infty dr\,r^4 R_{2n-1}(r) = -d^{s.c.}_p\,,
\end{equation}
with $ g_{v^n}=-\kappa[k(z) \partial_z\psi_{2n-1}(z)]^{z\to\infty}_{z\to-\infty}$ accounting for the vector meson contributions to the dipole. The known proportionality of the electric dipole moment to the dipole electromagnetic form factor at zero momentum, implies that the latter exhibits complete vector meson dominance. The same feature shows up for hadronic electromagnetic form factors in the CP-preserving sector of the WSS model \cite{SS2,SS-form}. 

The final result for the NEDM is obtained by taking into account the full neutron wave function\begin{equation}\label{nedmwf}
d_n = \frac{\int \rho^3 P(\rho)^2 \psi_Z^2(Z) d^{s.c.}_n d\rho\, dZ}{\int \rho^3 P(\rho)^2\psi_Z^2(Z) d\rho\,  dZ}\,.
\end{equation}

{\it 5. Results}. 
As a first result, note that the electric dipole moment of the proton, $d_p$, is the opposite of the neutron one. Numerically solving equation (\ref{eqforW}) we get 
\begin{equation}
d^{s.c.}_n = N_c \rho^4\frac{cm_q\pi}{\kappa M_{KK}^3}\theta D(\rho,Z) =  N_c\rho^4\frac{m_{\pi}^2}{M_{KK}^3}\theta D(\rho,Z)\,,
\label{leading}
\end{equation}
with $D(\rho_{cl},Z_{cl})\approx9.5\cdot 10^{-3}$ at $\lambda\gg1$. Notice the scaling with $N_c m_{\pi}^2$ which was also obtained in the Skyrme model approach \cite{dixon}. Then, we numerically compute  (\ref{nedmwf}), extrapolating the model parameters to real QCD data, setting $N_c=3$ and fixing, as customary, $\lambda = 16.63$, $M_\mathrm{KK} = 949 \,\mathrm{MeV}$ and $c m_q = 3.86\cdot 10^7\,\mathrm{MeV^4}$ from the matching of $f_{\pi}$, the rho meson mass and the GMOR relation. The final result is
\begin{equation}\label{full}
d_n = 1.8\cdot 10^{-16}\, \theta\;e\cdot \mathrm{cm}\,.
\end{equation}
It is interesting to estimate how much each vector meson mode contributes to the NEDM. 
Comparison of (\ref{leading}) with the numerical results obtained truncating the infinite sum in (\ref{nedmfinal}) to the first few modes, shows that retaining just the first vector mode overestimates the result of about 40\%. The contribution of higher modes is oscillating in sign. The inclusion of the first three modes is sufficient to obtain the complete result with per cent accuracy.

The (calculable) holographic model of (planar) QCD allows to properly take into account all the modes \footnote{A previous holographic calculation of the NEDM, in the simplest and less controllable bottom-up model (hard-wall), appears in \cite{hongNEDM}; the relation $d_n=-d_p$ appears in that model too.}. 
It is straightforward to show numerically that an increase of the coupling $\lambda$ enhances the importance of higher modes. The result we have obtained for the NEDM takes into account ``quantum" $1/N_c$ corrections included in the neutron wave function. 
Other $1/N_c$ and $1/\lambda$ corrections are expected to arise from terms which have been neglected in the model, but which could be consistently accounted for by its full string theory embedding. \\

{\it Acknowledgments.} -- We thank F. Becattini, C. Bonati, M. D'Elia, L. Martucci, H. Panagopoulos, D. Seminara, S. Sugimoto and E. Vicari for useful discussions. ALC is partly supported by the Florence Univ. grant ``Fisica dei plasmi relativistici: teoria e applicazioni moderne".\\

$^*$\small{Electronic addresses:~lorenzobartolini89@gmail.com, bigazzi@fi.infn.it, stefanobolo@gmail.com, cotrone@fi.infn.it, andrea.manenti@epfl.ch}

\end{document}